\documentstyle[12pt]{article}
\begin{document}


\vspace{.5cm}

\begin{center}
\Large{A METHOD FOR OBTAINING QUANTUM DOUBLES \\ FROM THE YANG-BAXTER
 $R$-MATRICES}
\end{center}

\vspace{.5cm}

\begin{center}
\large{A.A.VLADIMIROV}
\end{center}
\begin{center}
\large{Laboratory of Theoretical Physics, \\
Joint Institute for Nuclear Research, \\
Dubna, Moscow region, 141980, Russia}
\end{center}

\vspace{1cm}

\begin{center}
ABSTRACT
\end{center}

We develop the approach of refs.~\cite{FRT} and~\cite{MA} that
enables one to associate a quasitriangular Hopf algebra to
every regular invertible constant solution of the quantum
Yang-Baxter equations. We show that such a Hopf algebra is actually
a quantum double.

\vspace{7cm}

\_\_\_\_\_\_\_\_\_\_\_\_\_\_\_\_\_\_\_\_\_\_\_\_\_\_\_\_\_\_\_\_\_\_\_\_

E-mail: alvladim@theor.jinrc.dubna.su

\pagebreak

{\bf 1}. It is well known~\cite{FRT}
that any invertible constant matrix solution
$R$ of the quantum Yang-Baxter equation (QYBE)
\begin{equation}
R_{12}R_{13}R_{23}=R_{23}R_{13}R_{12}
\end{equation}
naturally generates a bialgebra $A_R=\{1,t_{ij}\}$
defined by
\[R_{12}T_1T_2=T_2T_1R_{12},\ \ \ \ \Delta (T_1)=T_1\otimes T_1,
 \ \ \ \ \varepsilon (T)={\bf 1},\]
(generators $t_{ij}$ form a matrix $T$, \ $\Delta$ is a coproduct
 and $\varepsilon$ a counit)
and also another bialgebra $U_R=\{1,l^+_{ij},l^-_{ij}\}$
with
\begin{eqnarray}
R_{12}L^{\pm}_2L^{\pm}_1 & = & L^{\pm}_1L^{\pm}_2R_{12}, \\
R_{12}L^+_2L^-_1  & = & L_1^-L^+_2R_{12}, \\
\Delta(L^{\pm}_1)  =  L^{\pm}_1\otimes L^{\pm}_1 &  , &  \ \
\varepsilon(L^{\pm}) = {\bf 1} ,
\end{eqnarray}
which is paired to $A_R$. This pairing~\cite{FRT,MA} is established
by the relations
$$ <T_1,L_2^+>=R_{12}, \ \ \ <T_1,L_2^->=R_{21}^{-1}, $$
obeys the duality conditions
$$ <\alpha\beta,a>=<\alpha\otimes\beta,\Delta(a)>,\ \ \
<\Delta(\alpha),a\otimes b>=<\alpha,ab>,$$
and appears to be degenerate.
 With some additional effort (quotienting
by appropriate null bi-ideals) these bialgebras can be made Hopf
algebras $\check{A}_R$ and $\check{U}_R$, dual to each other.
Their antipodes are defined by
\begin{eqnarray*}
<T_1,S(L_2^+)>&=&<S(T_1),L_2^+>=R_{12}^{-1}, \\
{}~<T_1,S(L_2^-)>&=&<S(T_1),L_2^->=R_{21}.
\end{eqnarray*}

With essential use of this duality Majid \cite{MA} showed that
in fact, with a certain reservation, $\check{U}_R$ proves to be
a quasitriangular Hopf algebra with the universal $\cal R$-matrix
 given by implicit formulas originated from
$<T_1\otimes T_2,{\cal R}>=R_{12}.$ By the way, Majid claims
{}~\cite{MA} that $\check{U}_R$ is `more or less' of the form of
a quantum double. In the present note we argue that, modulo the same
reservation, $\check{U}_R$ is actually a quantum double.

\vspace{.5cm}

{\bf 2}. Recall that a quantum double $ {\cal D}(A)$ is the Hopf
algebra of the following type (\cite{DR}, see also~\cite{FRT,MA,BU}).
 Let $A\otimes A^{\circ}$
be the tensor product of the Hopf algebra $A$ and its antidual
$A^{\circ}$.  Antiduality (i.e.\ the duality with opposite coproduct and
inverse antipode) means $<e^i,e_j>=\delta^i_j$ and \begin{equation}
<\alpha \beta,a> = <\alpha \otimes \beta,\Delta(a)>, \ \ <\Delta(\alpha),
a\otimes b>=<\alpha,ba>,
\end{equation}
$$ \varepsilon(a)=<1,a>,\ \ \varepsilon(\alpha)=<\alpha,1>,\ \
 <S(\alpha),a>=<\alpha,S^{-1}(a)> ,$$
where $a,b\in A,\ \alpha,\beta\in A^{\circ}$, and $\{e_j\},\{e^i\}$ are
the corresponding bases. To equip $A\otimes A^{\circ}$ with the Hopf
algebra structure of the quantum double, one must define a very specific
cross-multiplication recipe. If
$$ e_ie_j=c^k_{ij}e_k,\ \  \Delta(e_i)=f_i^{jk}(e_j\otimes e_k),\ \
S(e^i)=\sigma_j^ie^j, $$
it reads
$$ e^ie_j={\cal O}_{jq}^{ip}\,e_pe^q, \ \ \ \ \mbox{where}\
{\cal O}_{jq}^{ip}=c^t_{nq}c^i_{ts}\sigma_r^sf_j^{rl}f_l^{pn}. $$
In invariant form this looks like
\begin{equation}
\alpha a=\sum \sum<S(\alpha_{(1)}),a_{(1)}><\alpha_{(3)},a_{(3)}>
a_{(2)}\alpha_{(2)},
\end{equation}
where
$$\Delta^2(\alpha)=\sum \alpha_{(1)}\otimes \alpha_{(2)}\otimes\alpha_{(3)},
\ \ \ \Delta^2(a)=\sum a_{(1)}\otimes a_{(2)}\otimes a_{(3)}.$$
Here the usual notation for coproducts (cf.
$ \Delta(a)=\sum a_{(1)}\otimes a_{(2)} $ ) is used.
The resulting Hopf algebra proves to be quasitriangular
with the universal $\cal R$-matrix
$$ {\cal R}=\sum_{i}^{}(e_i\otimes1)\otimes(1\otimes e^i). $$

One easily finds that $gl_q(2)$ and other simple examples of quantum
universal enveloping algebras are both the $\check{U}_R$-type algebras
and quantum doubles. Can it happen that $\check{U}_R$ would be a
quantum double for any $R$? Majid's approach based on the $\check{U}_R
\leftrightarrow \check{A}_R$ duality does not readily
 answer this question.
That is why we choose another way: not to use $A_R$ at all. The key
observation is that there exists an inherent antiduality between
$U^+_R$ and $U^-_R$ which is precisely of the form required for the
quantum-double construction.

\vspace{.5cm}

{\bf 3}. Let us define bialgebras $U^+_R=\{1,l^+_{ij}\}$ and
$U^-_R=\{1,l^-_{ij}\}$ by eqs. (2),(4). Note that the
cross-multiplication relation (3) is not yet imposed, so $U^+_R$ and
$U_R^-$ are considered to be independent so far. However, the very
natural pairing between them can be introduced. It is generated by
\begin{equation}
<L_1^-,L^+_2>=R^{-1}_{12}, ~~~<L^-,{\bf 1}>=<{\bf 1},L^+>=<{\bf 1},{\bf 1}>=1
\end{equation}
and in the general case looks like
\begin{equation}
\left\langle L^-_{i_1}\ldots L^-_{i_m},L^+_{j_1}\ldots L^+_{j_n}\right\rangle
=R^{-1}_{i_1j_n}\ldots R^{-1}_{i_qj_p}\ldots R^{-1}_{i_mj_1} \ ,
\end{equation}
where the r.h.s.\  is a product of $mn \ \ R^{-1}$-matrices corresponding
to all pairs of indices $i_qj_p$ with $j$-indices ordered from right to left.
The consistency of (8) and (5) with (4) is evident,
while the proof of the consistency with (2) reduces to manipulations like
\begin{eqnarray*}
<L_0^-,R_{12}L^+_2L^+_1-L^+_1L^+_2R_{12}>&=&<L^-_0\otimes L_0^-,R_{12}(L
^+_1\otimes L^+_2)-(L^+_2\otimes L^+_1)R_{12}> \\
&=& R_{12}R^{-1}_{01}R^{-1}_{02}-R^{-1}_{02}R^{-1}_{01}R_{12}=0
\end{eqnarray*}
and repeated use of QYBE (1).

For general $R$, this pairing is degenerate. To remove the degeneracy,
i.e.\  to transform pairing into antiduality, one should factor out
appropriate bi-ideals \cite{MA}. In simple cases this procedure is
explicitly carried out and works well. For general $R$ it is of course
not under our control. The situation is quite similar to \cite{MA}: we
are to rely on that the factorization procedure is ``soft'' in a sense
that it does not destroy the whole construction.

\vspace{.5cm}

{\bf 4}. Keeping this in mind, we observe that, being antidual,
$U_R^{\pm}$ admit the Hopf algebra structure. Let us introduce an antipode
$S$ in $U_R^-$ and an inverse antipode $S^{-1}$ in $U^+_R$ by the relations
\begin{equation}
<S(L^-),{\bf 1}>=<{\bf 1},S^{-1}(L^+)>={\bf 1},\
<S(L_1^-),L^+_2>=<L_1^-,S^{-1}(L^+_2)>=R_{12},
\end{equation}
extending them on the whole of $U^+_R$ (or $U_R^-$) as
antihomomorphisms of algebras and coalgebras.  The definition is
correct due to \begin{eqnarray*} <m\circ(S\otimes
id)\circ\Delta(L^-_1),L^+_2>&=&<m\circ(S(L^-_1)\otimes L^-_1),L^+_2>\\
=<S(L_1^-)\otimes L_1^-,L^+_2\otimes L^+_2>&=&R_{12}R_{12}^{-1} ={\bf
1}=<\varepsilon(L^-_1),L^+_2>, \end{eqnarray*} $$
<S(L_0^-),R_{12}L^+_2L^+_1-L^+_1L^+_2R_{12}>= <\Delta\circ
S(L_0^-),R_{12}(L^+_1\otimes L^+_2)-(L^+_2 \otimes L^+_1)R_{12}> $$ $$
=<S(L_0^-)\otimes S(L_0^-),R_{12}(L_2^+\otimes L^+_1) -(L^+_1\otimes
L^+_2)R_{12}>=R_{12}R_{02}R_{01}-R_{01}R_{02}R_{12}=0,$$ $$
<S(R_{12}L_2^-L^-_1-L_1^-L_2^-R_{12}),L^+_0>=<R_{12}
S(L_1^-)S(L_2^-)-S(L_2^-)S(L_1^-)R_{12},L^+_0> $$
$$ =<\!R_{12}(S(L_1^-)\otimes S(L_2^-))-(S(L_2^-)\otimes
S(L_1^-))R_{12}, L^+_0\otimes
L^+_0\!>=R_{12}R_{10}R_{20}-R_{20}R_{10}R_{12}=0, $$ but there is no
such a formula for $S(L^+)$ or $S^{-1}(L^-)$. Once again the
factorization is hoped to be soft enough to allow the antipodes to be
invertible.

If so, our bialgebras $U_R^{\pm}$ become the mutually antidual Hopf
algebras $\check{U}_R^{\pm}$, and it is possible to define the
multiplicative structure of the quantum double upon
$\check{U}_R^+\otimes\check{U}_R^-$. The cross-multiplication rule is
deduced from (6):
\begin{eqnarray}
L_1^-L^+_2R_{12}&=&<S(L_1^-),L^+_2>L^+_2L_1^-<L_1^-,L^+_2>R_{12}\nonumber\\
&=& R_{12}L^+_2L_1^-R^{-1}_{12}R_{12}=R_{12}L^+_2L_1^-.
\end{eqnarray}
Thus we regain eq.(3) as the quantum-double cross-multiplication condition!

Our conclusion is that $R$-matrices obeying QYBE generate
 the algebraic structures of quantum double in quite a natural way.

\vspace{.5cm}

{\bf 5}. To illustrate the proposed scheme, consider the $sl_q(2)\ \
R$-matrix $$R_q=q^{-1/2}\left( \begin{array}{cccc}
q&0&0&0\\0&1&q-q^{-1}&0\\0&0&1&0\\0&0&0&q\end{array} \right),\ \ \   R_q^{-1}
=R_{q^{-1}}.$$
Here the bialgebras $U^{\pm}_{R_q}$ have 8 generators $l_{ij}^{\pm}$. The
bi-ideals to be factored out are generated by the relations
$$l^-_{21}=0,\ \  l^+_{12}=0,\ \  l^-_{11}l^-_{22}=l^-_{22}l^-_{11}=1,\ \
l^+_{11}l^+_{22}=l^+_{22}l^+_{11}=1.$$
After factorization the number of independent generators is reduced to
4.  We denote them $X^{\pm},H,H^{\prime}$ (note that $H^{\prime}\neq H$
so far):  $$L^+=\left( \begin{array}{cc}
q^{H/2}&0\\(q^{1/2}-q^{-3/2})X^+&q^{-H/2} \end{array} \right), \ \ \ \
L^-=\left( \begin{array}{cc}
q^{-H^{\prime}/2}&(q^{-1/2}-q^{3/2})X^-\\0&q^{H^{\prime}/2}
\end{array} \right).$$
The multiplication rules (inside each algebra),
coproducts and antipodes are:
\begin{eqnarray*}
[H,X^+]=2X^+&,&[H^{\prime},X^-]=-2X^-,\\
\Delta(H)=H\otimes1+1\otimes H&,&\Delta(H')=H'\otimes1+1\otimes H',\\
\Delta(X^+)=X^+\otimes q^{H/2}+q^{-H/2}\otimes X^+&,&
\Delta(X^-)=X^-\otimes q^{H'/2}+q^{-H'/2}\otimes X^-,\\
S(X^{\pm})=-q^{\pm 1}X^{\pm}&,&S(H)=-H,\ S(H^{\prime})=-H^{\prime}.
\end{eqnarray*}
The quantum-double cross-multiplication rules (6) take the form
\begin{eqnarray*}
[H^{\prime},X^+]=2X^+&,& [H,X^-]=-2X^-,\ \  [H,H^{\prime}]=0, \\
\ [ X^+,X^-] &=& \left( q^{(H+H^{\prime})/2}
-q^{-(H+H^{\prime})/2} \right)/(q-q^{-1}).
\end{eqnarray*}
The identification $H^{\prime}\equiv H$ leads to the ordinary $sl_q(2)$.

\vspace{.5cm}

{\bf 6}. To give one more illustration, let us consider a bialgebra
introduced in \cite{LU}. In a slightly simplified form it has
generators $\{1,t^i_j,u^i_j,E_j,F^i\}$ which obey the following
relations (here we prefer to display all the indices):
\begin{eqnarray}
R^{ij}_{mn}\,t^m_p\,t^n_q=R^{mn}_{pq}\,t^j_n\,t^i_m &,&
E_p\,t^j_q=R^{mn}_{pq}\,t^j_n\,E_m,\\
\Delta(t^i_j)=t^i_k\otimes t^k_j,\ \ \varepsilon(t^i_j)=\delta^i_j&,&
\Delta(E_j)=E_i\otimes t^i_j+1\otimes E_j,\ \ \varepsilon(E_j)=0,\\
R^{ij}_{mn}\,u^m_p\,u^n_q=R^{mn}_{pq}\,u^j_n\,u^i_m &,&
F^i\,u^j_p=R^{ji}_{mn}\,u^m_p\,F^n,\\
\Delta(u^i_j)=u^i_k\otimes u^k_j,\ \ \varepsilon(u^i_j)=\delta^i_j&,&
\Delta(F^i)=F^i\otimes1+u^i_j\otimes F^j,\ \ \varepsilon(F^i)=0,\\
R^{ij}_{mn}\,u^m_p\,t^n_q=R^{mn}_{pq}\,t^j_n\,u^i_m &,&
E_jF^i-F^iE_j=t^i_j-u^i_j,\\
u^i_p\,E_q=R^{mn}_{pq}\,E_n\,u^i_m&,&
t^i_p\,F^j=R^{ji}_{mn}\,F^m\,t^n_p,
\end{eqnarray}
with $R$ obeying QYBE~(1). This is not a bialgebra of the form
(2)-(4). Rather it is of the `inhomogeneous quantum group' type
{}~\cite{SC}. Let us make sure that it is a quantum double as well.

Consider $T,E$-bialgebra (11),(12) and $U,F$-bialgebra (13),(14)
firstly as being independent and fix nonzero pairings on the generators
by
\begin{equation}
<u^i_j,t^p_q\,>=R^{ip}_{jq},\ \ \ \
<u^i_j,1>=<1,t^i_j>=<F^i,E_j>=\delta^i_j,
\end{equation}
extending them to the whole bialgebras with the help of~(5).
The definition is correct due to
\begin{eqnarray*}
<F^i,E_p\,t^j_q-R^{mn}_{pq}\,t^j_n\,E_m>&=&
<F^i\otimes1+u^i_k\otimes F^k\ ,\ t^j_q\otimes E_p-R^{mn}_{pq}
(E_m\otimes t^j_n)> \\
&=&R^{ij}_{kq}\,\delta^k_p-\delta^i_mR^{mn}_{pq}\,\delta^j_n=0,\\
<F^iu^j_p-R^{ji}_{mn}\,u^m_p\,F^n,E_q>&=&
<F^i\otimes u^j_p-R^{ji}_{mn}(u^m_p\otimes F^n)\ ,\
E_k\otimes t^k_q+1\otimes E_q> \\
&=&\delta^i_kR^{jk}_{pq}-R^{ji}_{mn}\,\delta^m_p\,\delta^n_q=0.
\end{eqnarray*}

After factoring out the corresponding null bi-ideals, we may define
antipodes on the generators as follows:
$$ <S(u^i_j),t^p_q>=<u^i_j,S^{-1}(t^p_q)>={(R^{-1})}^{ip}_{jq}, $$
$$ <S(u^i_j),1>=<1,S^{-1}(t^i_j)>=\delta^i_j,\ \ S(F^i)=-S(u^i_j)F^j,\ \
S(E_j)=-E_iS(t^i_j). $$
The proof of correctness is in complete analogy with the
$\check{U}_R$-case.

Now a direct application of the recipe (6) exactly reproduces the
cross-multiplication relations (15),(16). For example,
$$ F^iE_j=<S(F^i),E_m><1,t^n_j>t^m_n+<S(u^i_n),1><1,t^m_j>E_mF^n $$
$$ +<S(u^i_n),1><F^m,E_j>u^n_m=-t^i_j+E_jF^i+u^i_j, $$
because of
$$ <S(F^i),E_m>=-<S(u^i_k)\otimes F^k\ ,\ E_n\otimes t^n_m+1\otimes
E_m>=-\delta^i_m. $$
Therefore, bialgebras of the
type (11)-(14) are also transformed into the quantum double using our
method.

\vspace{.5cm}

{\bf7}. Consider at last a bialgebra~\cite{BE} that is known to be
related~\cite{CA} to bicovariant differential calculus on quantum
groups.  Its coalgebra structure is given by (12), whereas the
multiplication relations (11) are to be supplemented by
\begin{eqnarray}
t^i_p\,E_q+f^i_{nm}\,t^n_p\,t^m_q&=&R^{nm}_{pq}\,E_m\,t^i_n+
f^n_{pq}\,t^i_n,\\
E_iE_j-R^{mn}_{ij}\,E_nE_m&=&f^m_{ij}\,E_m,
\end{eqnarray}
$f^i_{jk}$ being new structure constants. This bialgebra,
unlike its ancestor (11),(12), exhibits the $R$-matrix-type
representation
\begin{equation}
{\bf R}_{12}{\bf T}_1{\bf T}_2={\bf T}_2{\bf T}_1{\bf R}_{12},\ \
\Delta({\bf T})={\bf T}\otimes{\bf T},
\end{equation}
where, in terms of multi-indices like $I=\{0,i\}$,
$$ {\bf T}^I_J=\left( \begin{array}{cc}1&E_j\\0&t^i_j\end{array}
\right),\ \ \ {\bf R}^{IJ}_{MN}=\left( \begin{array}{cccc}
1&0&0&0\\0&\delta^j_n&0&0\\0&0&\delta^i_m&f^i_{mn}\\0&0&0&R^{ij}_{mn}
\end{array} \right). $$
Of course, ${\bf R}$ must satisfy the QYBE (1) which now involves
the structure constants $f^i_{mn}$ as well as $R^{ij}_{mn}$. Note that,
due to (18),(19), the bialgebra (11),(12) is not restored from (20)
by mere setting $f^i_{mn}\equiv0$.

Now let us try to develop a quantum double from the bialgebra (20).
However, it seems to be quite uneasy task. A natural Ansatz for the
candidate antidual bialgebra is
$$ {\bf U}^I_J=\left( \begin{array}{cc}
1&0\\F^i&u^i_j \end{array} \right), $$
which causes the corresponding $R$-matrix to be
$$ {\bf\overline{R}}^{IJ}_{MN}=\left( \begin{array}{cccc}
1&0&0&0\\0&\delta^j_n&0&0\\0&0&\delta^i_m&0\\0&\bar{f}_n^{ij}&0&
R^{ij}_{mn} \end{array} \right) $$
with different structure constants $\bar{f}$ and another QYBE system
involving $R$ and $\bar{f}$. Now, attempting to
fix a pairing in the form
\begin{equation} <{\bf U}_1,{\bf T}_2>={\bf Q}_{12}
\end{equation}
with a certain numerical matrix ${\bf Q}$, we immediately
arrive at the following general statement:

Let ${\bf R}$ and ${\bf\overline{R}}$ be invertible solutions of
QYBE.  If there exists an invertible solution ${\bf Q}$ of the
equations
\begin{eqnarray*}
{\bf Q}_{12}{\bf Q}_{13}{\bf R}_{23}&
 =&{\bf R}_{23}{\bf Q}_{13}{\bf Q}_{12},\\
{\bf\overline{R}}_{12}{\bf Q}_{13}{\bf Q}_{23}&
=&{\bf Q}_{23}{\bf Q}_{13}{\bf\overline{R}}_{12},
\end{eqnarray*}
then (21) is a correct pairing between the ${\bf T}$-~and
${\bf U}$-bialgebras generated by ${\bf R}$ and
$\bf\overline{R}$, respectively, and, assuming a proper
quotienting procedure to be performed, the antipodes can be defined by
the relations
$$<S({\bf U}_1),{\bf T}_2>=<{\bf U}_1,S^{-1}({\bf T}_2)>=
{\bf Q}^{-1}_{12} $$
and the quantum-double structure can be established on the tensor
product of these bialgebras by the cross-multiplication formula
$$ {\bf Q}_{12}{\bf U}_1{\bf T}_2={\bf T}_2{\bf U}_1
{\bf Q}_{12}.$$

Whether such a program can really be carried through in interesting
cases (e.g. for $\bf R$ and $\bf\overline{R}$ given above) is the
subject of further investigation.

\vspace{.5cm}

I wish to thank A.Isaev, R.Kashaev, A.Kempf and P.Pyatov for
stimulating discussions.

\end{document}